\documentclass[a4paper]{article}

\usepackage[left=3cm, right=3cm, top=2cm, bottom = 2cm]{geometry}

\usepackage[utf8]{inputenc}
\usepackage[T1]{fontenc}
\usepackage{lmodern}
\usepackage[UKenglish]{isodate}

\usepackage[perpage]{footmisc}

\usepackage{amsmath}
\usepackage{graphicx}
\usepackage{xcolor}
\usepackage{tikz}
\usepackage{textcomp}

\usepackage{url}
\usepackage{cite}

\usepackage{scalerel}
\usepackage{tikz}
\usetikzlibrary{svg.path}

\definecolor{orcidlogocol}{HTML}{A6CE39}
\tikzset{
  orcidlogo/.pic={
    \fill[orcidlogocol] svg{M256,128c0,70.7-57.3,128-128,128C57.3,256,0,198.7,0,128C0,57.3,57.3,0,128,0C198.7,0,256,57.3,256,128z};
    \fill[white] svg{M86.3,186.2H70.9V79.1h15.4v48.4V186.2z}
                 svg{M108.9,79.1h41.6c39.6,0,57,28.3,57,53.6c0,27.5-21.5,53.6-56.8,53.6h-41.8V79.1z M124.3,172.4h24.5c34.9,0,42.9-26.5,42.9-39.7c0-21.5-13.7-39.7-43.7-39.7h-23.7V172.4z}
                 svg{M88.7,56.8c0,5.5-4.5,10.1-10.1,10.1c-5.6,0-10.1-4.6-10.1-10.1c0-5.6,4.5-10.1,10.1-10.1C84.2,46.7,88.7,51.3,88.7,56.8z};
  }
}

\newcommand\orcidicon[1]{\href{https://orcid.org/#1}{\mbox{\scalerel*{
\begin{tikzpicture}[yscale=-1,transform shape]
\pic{orcidlogo};
\end{tikzpicture}
}{|}}}}

\usepackage{paralist}
\usepackage{nameref}

\usepackage[hidelinks,breaklinks=true ]{hyperref}
\usepackage[strings]{underscore}
\usepackage{breakurl}

\usepackage[xindy]{glossaries}

\newacronym{vnv}{V\&V}{Verification and Validation}

\newacronym{asil}{ASIL}{Automotive Safety and Integrity Level}
\newacronym{dal}{DAL}{Design Assurance Levels or Development Assurance Level}

\newacronym{mcs}{MCS}{Mixed-Criticality System}

\newacronym{fm}{FM}{Formal Methods}
\newacronym{as}{AS}{Autonomous System}

\newcommand{\Uppaal}{\textsc{Uppaal}}

\newacronym{rv}{RV}{Runtime Verification}

\newacronym{rml}{RML}{Runtime Monitoring Language}

\newacronym{ltl}{LTL}{Linear-time Temporal Logic}

\newacronym{ctl}{CTL}{Computation Tree Logic}

\newacronym{ptl}{PTL}{Probabilistic Temporal Logic}

\newacronym{pctl}{PCTL}{Probabilistic Computation Tree Logic}

\newacronym{pctl*}{PCTL*}{Probabilistic Computation Tree Logic*}

\newacronym{tctl}{TCTL}{Timed Computation Tree Logic}

\newacronym{stl}{STL}{Signal Temporal Logic}

\newacronym{ltl-x}{LTL-X}{a version of Linear Time Logic without the `next' operator}

\newacronym{mtl}{MTL}{Metric Temporal Logic}

\newacronym{iactlk-x}{IACTLK-X}{a fragment of the temporal-epistemic logic CTLK, without the `next' operator}

\newacronym{mucalc}{$\mu$-calculus}{a strict superset of other temporal logics}

\newacronym{fotl}{FOTL}{First-Order Temporal Logic}

\newacronym{bdi}{BDI}{Belief-Desire-Intention}

\newacronym{prs}{PRS}{Procedural Reasoning System}

\newacronym{are}{ARE}{Autonomous Reasoning Engine}

\newacronym{dmars}{dMARS}{distributed Multi-Agent Reasoning System}

\newacronym{fsm}{FSM}{Finite-State Machine}

\newacronym{pfsm}{PFSM}{Probabilistic Finite-State Machine}

\newacronym{apfsm}{APFSM}{\textit{Autonomous} Probabilistic Finite-State Machine}

\newacronym{pha}{PHA}{Probabilistic Hybrid Automata}

\newacronym{fsa}{FSA}{Finite-State Automata}

\newacronym{ta}{TA}{Timed Automata}

\newacronym{gspn}{GSPN}{Generalised Stochastic Petri Net}

\newacronym{mopn}{MOPN}{Marked Ordinary Petri Net}

\newacronym{ba}{BA}{B\"{u}chi Automata}

\newacronym{dtmc}{DTMC}{Discrete-Time Markov Chain}

\newacronym{pars}{PARS}{Process Algebra for Robot Schemas}

\newacronym{fsp}{FSP}{Finite State Processes}

\newacronym{piadl}{$\pi$ADL}{$\pi$ calculus combined with ADL}

\newacronym{csp}{CSP}{Communicating Sequential Processes}

\newacronym{ccs}{CCS}{Calculus of Communicating Systems}

\newacronym{wsccs}{WSCCS}{Weighted Synchronous CCS}

\newacronym{csp|b}{CSP$\|$B}{an integrated formalism combining CSP and B}

\newacronym{fdr}{FDR}{the Failures-Divergences Refinement checker}

\newacronym{jpf}{JPF}{Java PathFinder}

\newacronym{ajpf}{AJPF}{Agent Java PathFinder}

\newacronym{mcmas}{MCMAS}{Model Checker for Multi-Agent Systems}

\newacronym{ltlmop}{LTLMoP}{Linear Temporal Logic MissiOn Planning}

\newacronym{fret}{FRET}{Formal Requirements Elicitation Tool}

\newacronym{adl}{ADL}{Architecture Description Language}

\newacronym{aadl}{AADL}{Architecture Analysis and Design Language}

\newacronym{lts}{LTS}{Labelled-Transition System}

\newacronym{uml}{UML}{Unified Modelling Language}

\newacronym{sysml}{SySML}{Systems Modelling Language}

\newacronym{robotml}{RobotML}{Robot Modelling Language}

\newacronym{dsl}{DSL}{Domain Specific Language}

\newacronym{sct}{SCT}{Supervisory Control Theory}

\newacronym{ide}{IDE}{Integrated Development Environment}

\newacronym{mde}{MDE}{Model-Driven Engineering}

\newacronym{ai}{AI}{Artificial Intelligence}

\newacronym{fdir}{FDIR}{Fault Detection, Isolation and Recovery}

\newacronym{ifm}{iFM}{Integrated Formal Methods}

\newacronym{ros}{ROS}{Robot Operating System}

\newacronym{dl}{\text{\upshape\textsf{d{\kern-0.05em}L}}}{differential dynamic logic}

\newacronym{vm}{VM}{Virtual Machine}

\newacronym{bip}{BIP}{Behaviour Interaction Priority}

\newacronym{fol}{FOL}{First-Order Logic}

\newacronym{owl}{OWL}{Web Ontology Language}

\newacronym{ria}{RIA}{Restore Invariant Approach}

\newacronym{ocl}{OCL}{Object Constraint Language}

\newacronym{sil}{SIL}{Safety Integrity Level}

\newacronym{rcl}{RCL}{ROS Contract Language}
\makeglossaries

\usepackage[disable]{todonotes}

\usepackage[pagewise]{lineno}

\begin{document}

\title{Using Formal Methods for Autonomous Systems: Five Recipes for Formal Verification\thanks{
Thanks are owed to Marie Farrell, for our previous (and current) collaborations and for her \textit{powerful} proof-reading; and Louise Dennis, Angelo Fernando, Rafael C. Cardoso, and Michael Fisher, for their helpful conversations about the work that went into this paper. Parts of this work were completed while the author was employed by the University of Liverpool and then the University of Manchester, UK.
Work supported through the UKRI Robotics and AI in Nuclear (RAIN) grant EP/R026084. }}

\vspace{2em}
\author{\vspace{2em}Matt Luckcuck \orcidicon{0000-0002-6444-9312} \\ \small{Department of Computer Science, University of Manchester, UK}}
\date{\today}

\maketitle 
\vspace{2em} 

\begin{abstract}
Formal Methods are mathematically-based techniques for software design and engineering, which enable the unambiguous description of and reasoning about a system’s behaviour.
Autonomous systems use software to make decisions without human control, are often embedded in a robotic system, are often safety-critical, and are increasingly being introduced into everyday settings.
 Autonomous systems need robust development and verification methods, but formal methods practitioners are often asked: \textit{Why use Formal Methods for Autonomous Systems?}
To answer this question, this position paper describes five recipes for formally verifying aspects of an autonomous system, collected from the literature. The recipes are examples of how Formal Methods can be an effective tool for the development and verification of autonomous systems.
During design, they enable unambiguous description of requirements; in development, formal specifications can be verified against requirements; software components may be synthesised from verified specifications; and behaviour can be monitored at runtime and compared to its original specification. 
Modern Formal Methods often include highly automated tool support, which enables exhaustive checking of a system's state space. 
This paper argues that Formal Methods are a powerful tool for the repertoire of development techniques for safe autonomous systems, alongside other robust software engineering techniques.

\end{abstract}

\maketitle 
\glsresetall

\vspace{2em}
\section{Introduction}
\label{sec:intro}

Autonomous systems use software to make decisions without the need for human control. They are often embedded in a robotic system, and are increasingly being introduced into situations where they are near to or interact (physically or otherwise) with humans. These features mean that autonomous systems -- and especially autonomous \textit{robotic} systems -- are often safety-critical, where failures can cause harm to humans or even their death.

The term \gls{fm} describes a wide range of mathematically-defined techniques for specifying how a system should behave or how it will operate on data, and robustly reasoning about these specifications. \gls{fm} are a useful tool for software engineering. Despite this, discussions of using \gls{fm} are often still dogged by myths, both supportive and critical, that have persisted for decades -- for example the `myths' papers~\cite{Hall1990,Bowen1995a}, which were published in 1990 and 1995, respectively. \gls{fm} are effective in real-world situations, as shown by their successful use in many industrial projects~\cite{Woodcock2009}; and \gls{fm} have been used in the literature to overcome the challenges faced when specifying and verifying autonomous systems~\cite{luckcuck_formal_2019}. But \gls{fm} researchers and practitioners are still often asked:\\
\centerline{\textit{Why use Formal Methods for Autonomous Systems?}}
\newpage
\noindent In this position paper, we answer this question by:~
\begin{itemize}
\item describing four approaches to using \gls{fm} for engineering safe autonomous systems, and
\item presenting five \textit{recipes} for the formal verification of components or aspects of an autonomous system.
\end{itemize} 
Throughout the paper, we argue that \gls{fm} are useful to have in one's repertoire when verifying autonomous systems. This paper often discusses autonomous systems that control a robotic system, because of the extra safety requirements this brings. However, the majority of the recipes that we present can be applied to general autonomous systems.

Autonomous systems need to be safe, correct, and trustworthy, so the most robust design and verification methods available should be used to reduce the risks to humans to as low as is reasonably practicable. For autonomous systems used in industry, the people at risk are likely to be workers; for systems like autonomous vehicles and domestic assistants, the people at risk will be the system's users (for example the people in the autonomous car or the person receiving domestic assistance) and bystanders (for example pedestrians or household visitors). Security flaws in autonomous systems are also a concern, both because of the sensitive data they are likely to contain and because a security failure can cause a safety failure. Verifying that the autonomous system preserves these safety and security requirements is a crucial activity in the systems development life cycle.

According the Software Engineering Body of Knowledge (SWEBOK), version 3\cite{swebok} ``Verification is an attempt to ensure that the product is built correctly, in the sense that the output products of an activity meet the specifications imposed on them in previous activities.'' Whereas ``Validation is an attempt to ensure that the right product is built -- that is, the product fulfils its specific intended purpose.'' Or, in the informal description of Boehm~\cite{boehm_verifying_1984}, verification answers the question ``Am I building the product right?'', and validation answers the question ``Am I building the right product?''. While both of these activities are important, this paper discusses `verification', in the sense of evaluating a system's compliance with a specification of its requirements, and uses the term `formal verification' to mean verification by \gls{fm}.

As part of previous work, we identified the challenges that are faced when formally specifying and verifying the behaviour of (autonomous) robotic systems~\cite{luckcuck_formal_2019}. To summarise~\cite{luckcuck_summary_2019}, these challenges can be \textit{external} or \textit{internal} to the autonomous system. The two external challenges were that of modelling the system's operating environment, and of providing evidence that the public or a regulator should trust the system. Arguably, the biggest external challenge is the system's operating environment. This is where the \textit{reality gap}, the difference between design-time models of the environment and the real world, can cause an autonomous system to fail. 
The internal challenges were related to how the autonomy was implemented (which we discuss in detail in the \nameref{sec:autonomousSystems} section) and how the system's architecture was arranged. Arguably, the internal challenge that is most important to overcome is that of checking that an autonomous system's executive decisions are correct.

\gls{fm} can be applied across the systems development life cycle, from requirements through to operation. Formal specifications (and sometimes software code) can be formally verified as conforming to a set of properties -- a simple example is the property of deadlock freedom. Like verification by testing or simulation, formal verification requires careful thought about the level of abstraction used to represent the system and its environment, and about what properties to check the system for. This takes a level of skill that is similar to producing good tests or good simulations. 

Each of our formal verification recipes is drawn from examples in the literature, but the recipes are described in a generic way so as to encourage their application to a wide range of autonomous systems. The recipes help to fill the gap in guidance for how to apply \gls{fm} to autonomous system in a systematic way. 
A recipe is implemented by using a formal \textit{approach}, by which we mean the different ways of using a formal notation such as model checking or runtime verification. Thus, our recipes use formal approaches, which themselves use formal notations or languages. 

The formal approaches described in this paper, and the way that we use the term `approach' here, are drawn from our previous survey paper~\cite{luckcuck_formal_2019}. The recipes are drawn from the same survey and from research experience during the Robotics and AI in Nuclear (RAIN) project, one of the UK’s \textit{Robots for a Safer World}\footnote{\url{https://www.ukri.org/our-work/our-main-funds/industrial-strategy-challenge-fund/future-of-mobility/robots-for-a-safer-world-challenge}} research and innovation hubs; some of this previous experience is summarised in~\cite{fisher_overview_2021}.

The rest of this paper is organised as follows. The second section, \nameref{sec:autonomousSystems}, introduces the basic concepts of autonomous systems and explains the challenges they pose to verification efforts. The third section, ~\nameref{sec:approaches}, introduces four popular approaches to using formal methods. In the fourth section \nameref{sec:verificationTactics}, we describe the five recipes for using \gls{fm} for the verification of specific aspects or components of an autonomous system. Finally, the fifth section, \nameref{sec:conclusion}, concludes the paper.

\section{Autonomous Systems}
\label{sec:autonomousSystems}

Autonomous systems should be designed with verification in mind, especially where they are 
mission-critical\footnote{A mission-critical system is one where a failure may lead to large data- or financial-losses.} or safety-critical\footnote{A safety-critical system is one where a failure may lead to ecological or financial disaster, serious injury, or death.} systems. Autonomous systems combine software that makes decisions and interprets sensor signals, with hardware that physically interacts with the real-world. Both of these elements pose challenges for verification.

Autonomy can be implemented using, broadly, two different styles of \gls{ai}: symbolic and sub-symbolic. As discussed in the \nameref{sec:TypesofAutonomy} section, symbolic approaches tend to be easier to inspect and formally verify. 

An autonomous system may contain components implemented in each of these styles, and it is important to choose the right \gls{ai} technique for the right function. The criteria for choosing which \gls{ai} technique to use are similar to when choosing a human to do the same job: how can its decisions be checked for correctness, and how confident are we in the checks that are available. This is especially important when assurance evidence is needed for regulatory approval. 

The introduction of robotic and autonomous systems can raise a variety of ethical issues, which may not be safety or security hazards but still have a negative impact on user trust and societal acceptability. 
Current ethics guidance for \gls{ai} is fragmented, but a recent survey~\cite{Jobin2019} found convergence on five principles: transparency, justice and fairness, non-maleficence, responsibility, and privacy, with a further six principles: beneficence, freedom and autonomy, trust, sustainability, dignity and solidarity mentioned in many guideline documents. Further, the IEEE are currently developing the P7000 series of standards, which cover a variety of aspects of ethics for autonomous systems\footnote{IEEE P7000 standards: \url{https://ethicsinaction.ieee.org/p7000/}}; including: P7001 `Transparency of Autonomous Systems', P7003 `Algorithmic Bias Considerations', and P7009 `Fail-Safe Design of Autonomous and Semi-Autonomous Systems'.

The BS~8611 standard \emph{`ethical design and application of robots and robotic systems'}~\cite{EthicalDesignofRobots2016}  uses the term \emph{ethical hazards}: a potential source of ethical harm, which is anything likely to compromise psychological, societal, and environmental well-being. Any hazard assessment should include the assessment of ethical hazards, and they should be designed out of the final system. These design-time efforts could be paired with run-time techniques, for example, an \textit{ethical black box}~\cite{Winfield2017} that records the sensor input and internal state of a system to enable offline checking of the system's decisions in the event of a failure. Leaving ethical harms to be discovered or handled as part of the operation phase of the system is, by definition, unethical. Ethical harms that are discovered while the system is in-use can directly impact the safety or security of the system or cause huge reputational damage to the system's developers or operators, both of which can have financial impacts as well.

When interacting with the real world, an autonomous system will have to cope with some degree of uncertainty. Sensor input may be noisy and unreliable, and \gls{ai} techniques for recognising objects are often inherently statistical. Movement and navigation is usually aided by internal (often pre-built) models of the environment, but if the environment changes and the model isn't updated, then using it can become unreliable. 

These elements of autonomous systems can be a challenge for both formal and informal verification methods. Autonomous systems combine hardware and software components that use expertise from various sub-domains of computer science, electronics, and engineering. Each of these sub-domains often has their own domain-specific verification approaches, and therefore the most trusted or suitable verification approach might be different for different components. For this reason, our previous work argues that the verification of autonomous systems benefits from the integration of formal and informal verification techniques~\cite{Farrell2018}. The \nameref{sec:VerifyingAutonomy} section discusses in detail the challenges that autonomous systems pose to both formal and informal verification. 
 
\subsection{Types of Autonomy}
\label{sec:TypesofAutonomy}

There are broadly two \gls{ai} approaches used in implementing autonomous systems: symbolic and sub-symbolic. Symbolic approaches include logic programming (for example Prolog  ) and agent programming languages (for example, Jason\cite{Bordini07}), where statements and rules are used to describe the world and available actions. Because these systems are explicit and based on mathematical logic, they are (relatively) easier to inspect. However, engineering the knowledge they need (the statements and rules) can present an overhead. Also, problems that cannot be easily expressed as logical statements and rules are difficult to encode in symbolic systems. 

Sub-symbolic (or connectionist) approaches to \gls{ai} include machine learning and neural networks, where inferences are made about large amounts of existing data to build knowledge about the world, often using statistical methods. This can make these approaches better suited to situations where \textit{how} to reach a decision is not well understood or not easy to describe logically. However, these approaches are much more difficult to inspect, meaning that their decisions are less amenable to verification.

Clearly, the approach(es) chosen will impact the techniques available for verifying the system's decisions, and so impact the assurance argument. It is likely that an autonomous system might need both types of \gls{ai}, for example a symbolic approach to planning but a sub-symbolic approach to vision classification. Other work in the literature already combines logical symbolic approaches with flexible sub-symbolic approaches~\cite{Anderson2018, Aitken2018}. So it is useful to have verification techniques that can target both types of \gls{ai} approach.

Where an autonomous system will be used in safety-critical scenarios, using symbolic approaches for high-level decisions and planning enables robust verification of the system's choices, as well as providing a route to explaining those choices. Confining sub-symbolic approaches to functions to which they are better suited (functions that are often not amenable to symbolic approaches) reduces the number of components that are difficult to inspect. The system's architecture can then be used to prevent potential failures caused by a sub-symbolic component from propagating to the overall behaviour of the system. This is discussed in more detail in various subsections of the \nameref{sec:verificationTactics} section.

\subsection{Levels of Autonomy}
\label{sec:LevelsofAutonomy}

In addition to how an autonomous system is implemented, there are also various ways of describing how much, or what level, of autonomy a system has. For example, an autonomous robotic system might be described as semi-autonomous or fully autonomous. A semi-autonomous robotic system is one where the ``robot and a human operator plan and conduct the task, requiring various levels of human interaction''~\cite{IEEE1872-2015}. In a fully-autonomous robotic system, the robot performs its task ``without human intervention, while adapting to operations and environmental conditions''~\cite{IEEE1872-2015}.

There are also frameworks that describe the different \textit{levels} of autonomy that a system might have. The original levels of autonomy were defined for undersea teleoperation systems~\cite{sheridan_human_1978} -- though they are worded rather neutrally, so they may have wider applicability -- and were revised in~\cite{endsley_level_1999}. A detailed review of frameworks for levels of autonomy can be found in Beer et al.'s work on developing a framework for levels of autonomy in Human-Robot Interaction~\S3~of~\cite{beer_toward_2014}.

Some frameworks for levels of autonomy are generic and aimed at all autonomous systems, for example Huang et al.~\cite{huang_autonomy_2005}, who describe autonomy levels for `unmanned' systems generally; and Beer et al.~\cite{beer_toward_2014}, who describe levels of autonomy from a Human-Robot Interaction perspective. Other frameworks are specific to a particular sector or operational environment, for example spacecraft~\cite{ryan_proud_methods_2003} or autonomous cars~\cite{on-road_automated_driving_orad_committee_sae_international_taxonomy_2018}. Adapting or reusing a framework for levels of autonomy in a new context might seem attractive, but this can lead to confusion because the levels that are defined for one environment may not clearly map onto another.

Frameworks and definitions of the type or level of autonomy that a system possesses can be useful for describing its capabilities. However, there is no single framework that has been universally adopted, so it is important to be clear about \textit{which} framework or definition is being used.

\subsection{Challenges to Verification}
\label{sec:VerifyingAutonomy}


Autonomous systems pose challenges to verification, especially when they are embedded in a robotic system~\cite{Farrell2018}. As previously mentioned, by `verification' we specifically mean the evaluation of a system's compliance with a specification. The two key challenges when verifying autonomous systems are: that the system is making decisions for itself and it is operating in (and possibly physically interacting with) the real world. These challenges are faced by all types of verification approaches, but a detailed survey of the current challenges being tackled specifically in the \gls{fm} literature is presented in~\cite{luckcuck_formal_2019}.

As described in the \nameref{sec:TypesofAutonomy} section, the way that an autonomous component is implemented has a notable impact on what verification techniques can be applied to it. Symbolic \gls{ai} (such as agent-based systems) are often easier to verify, whereas Sub-Symbolic (connectionist) \gls{ai} can prove more difficult. Ideally, verifiability should be designed into the system; autonomous components should be verifiable, the number of autonomous components that are difficult to analyse should be limited, and careful mitigations should be introduced where difficult-to-analyse autonomous components are needed.  Mitigations could include adopting an architecture that defends the system from components with lower levels of assurance, using runtime software checks, or using \gls{fm} approaches. The \nameref{sec:verificationTactics} section presents five recipes, built from \gls{fm}, to mitigate these challenges by enabling the verification of different parts of an autonomous system.

Aside from the challenges that autonomous software presents, it also provides the opportunity to examine a system's decision-making mechanisms. This requires designing autonomous software with verification in mind, and selecting \gls{ai} techniques that can be inspected. Examining the system's decision-making is especially important when safety assurance evidence is required, or when the system's choices need to be compared with a qualified human operator. Ignoring the chance to interrogate the decisions that an autonomous system is making would be a wasted opportunity.


Where an autonomous system is embedded within a robotic system, the robotic parts are often checked with physical tests, but doing this systematically can be challenging. Physical tests with a robot can be dangerous to the human testers in the early phases of development, and are often difficult or time-consuming to set-up and run. This means that simulations and code analysis are helpful, even though these may lack the fidelity of the actual operating environment. Despite this, there is evidence that even a low-fidelity simulation of a robot's environment can reproduce bugs that were found during field tests of the same robot; Sotiropoulos et al.\cite{Sotiropoulos2017} found that of the 33 bugs that occurred during a field test of a robot, only one could not be reproduced in their low-fidelity simulation.

Informal verification techniques have been developed for some of the frequently used control and autonomy approaches. 
For example, there are well established checks to assess the correctness of feedback control mechanisms. Usually, feedback controllers model the system that they are controlling using differential equations. It is important to check that: (1) the model is validated against the real world, and that it is based on valid data; (2) any abstractions in the implementation of the feedback controller are valid, and; (3) the feedback controller is robust against variations in the system's performance, and can cope with any differences between its model and the real system.

Similarly, techniques are being developed for the verification and testing of Deep Neural Networks (DNNs) -- and a survey of the current state of the art can be found in~\cite{huang_survey_2019}. A key correctness property for a DNN is whether two inputs that appear identical to a human observer could be classified differently by the DNN. When safety assurance evidence is needed for a DNN, it is important to be clear about how the correctness properties are defined and how they have been assessed.

Without being specifically developed for autonomous systems, \gls{fm} can be applied across autonomous system's development lifecycle. 
In the requirements and design phases, \gls{fm} can provide an unambiguous specification of the system's requirements or its design. 

Requirements formalisation was the most often used formal technique in the industrial projects surveyed in~\cite{Woodcock2009}. 
A recent survey of the literature for formal techniques for capturing requirements can be found in~\cite{bruel_role_2021}, which also emphasises that a formal representation of requirements can support other representations, instead of just replacing them.

On the spectrum from natural-language requirements to fully-formalised requirements, the \gls{fret} sits in the middle, using a structured natural language that can be automatically translated into temporal logic~\cite{madhavji_generation_2020}. Even if not using the temporal logic translation, \gls{fret} can help to bridge the gap between informal requirements and assurance arguments, and properties in other formalisms, for example Event-B~\cite{bourbouh_integrating_2021}

Formal designs can be used to generate provably correct source code from specifications and designs. For example, translating high-level system designs written in \gls{aadl} into the \gls{bip} framework to facilitate controller generation~\cite{chkouri_translating_2008}, or the generation of movement plans for an autonomous robotic system~\cite{karaman2009sampling, kress2011correct}. This approach can be seen as similar to Model-Driven Engineering, but using formal models. This can help to minimise the likelihood of introducing faults during software development. Generating control programs from high-level formal specifications is described in more detail in the \nameref{sec:synthesis} section.

\gls{fm} can also be used to perform rigorous static and dynamic analysis. Static analysis can be performed on more abstract models of the system, but techniques exist that are capable of operating directly on program code itself -- for example the Bounded Model Checkers for C programs (CBMC\footnote{\url{http://www.cprover.org/cbmc/}}) and Java programs (JBMC\footnote{\url{http://www.cprover.org/jbmc/}}), and the Model Checking Agent Programming Language (MCAPL) framework\cite{Dennis2018} for agent programs. Using \gls{fm} here has the benefits of being automatic and exhaustive. \gls{fm} are capable of accepting models that talk about time and probabilities. For tackling very large specifications, there are also static analysis techniques that take a similar approach to statistical testing, taking samples of the available paths through a model. Two static analysis approaches are discussed in the \nameref{sec:modelChecking} and \nameref{sec:theoremProving} sections, respectively.

For dynamic formal analysis, \gls{fm} can monitor a running system, checking it against a formal specification of its expected behaviour. Effectively, this re-runs the design-time verification but with the behaviour of the system as input. If the system's behaviour differs from the specification, then the monitor can log the failure, alert the user, or trigger mitigating actions. This is described in more detail in the \nameref{sec:rv} section.

It is likely that some components in an autonomous system are more critical to the system's safe operation than others, this is referred to as a mixed-criticality system. The more critical a component is, the higher the assurance against failure that the component needs. It may be useful to analyse the criticality of the system's components and apply \gls{fm} to the most critical (where the component is amenable to \gls{fm}). This suggestion aims to  ensure that the most critical components are assured against failure, without requiring the entire system to be formally verified. For a comprehensive review of mixed-criticality research see\cite{burnsMixedCriticalitySystems2018}.

The connection between an autonomous system and its environment is often implemented in a general purpose programming language, which can make it more difficult to use formal verification techniques. Where an autonomous system is connecting to the real world, the \gls{ros}\footnote{\url{https://www.ros.org/}} middleware is often used. A \gls{ros} system is a network of communicating nodes, programmed in either C++ or Python. For autonomous systems linked to simulated worlds, the landscape is more fragmented but examples include an autonomous agent linked to the flight simulator FLIGHTLAB~\cite{Webster2014}, and a connector to link multi-agent systems to the video game StarCraft~\cite{koeman_designing_2019}. Using connectors or middleware can make programming the system easier, but it is another component of the system that could cause (or contribute to) failures.

In addition to the off-the-shelf \gls{fm}, there are also techniques that are specialised to autonomous systems, capable of tackling the challenges that they pose during verification~\cite{luckcuck_formal_2019}. In the next section, we introduce four approaches that use \gls{fm}, describe them in detail, and give examples of their use in the verification of autonomous systems.

\section{Formal Verification Approaches}
\label{sec:approaches}

This section presents an overview of four approaches to using \gls{fm}, often found in the literature. As previously mentioned, by an `approach' we mean \textit{how} a formal notation is used to verify properties. Each approach is illustrated by examples from the literature. This is nowhere near an exhaustive list, but is intended to show a range of approaches and introduce the reader to the approaches used in the recipes, which are described in the next section. 

In addition to applying single formal approaches, there is a field of research investigating the \textit{integration} of different formal (and informal) verification approaches -- for example, the integrated Formal Methods (iFM) conference series has been running since 1999~\cite{araki_ifm99_1999}. There is significant scope in the area of robotics and autonomous systems to use integrated approaches~\cite{Farrell2018, gleirscher_new_2020} because these systems combine components from many different disciplines. The \textit{Corroborative \gls{vnv}} approach ~\cite{websterCorroborative2020} is specifically designed to integrate formal and informal models of the same system to support wholistic verification. A fuller description of the formal approaches that have been applied to autonomous systems in recent literature, including integrated approaches, can be found in our previous survey paper~\cite{luckcuck_formal_2019}.

\subsection{Model-Checking} 
\label{sec:modelChecking}

Model Checking determines whether or not a property holds in every state of a formal specification. The specification and the property may be written in the same notation, but they are often not. For example, many model checking tools accept the specification as some form of \gls{fsm} with the property written in some form of temporal logic. 

Model checking is a flexible, exhaustive formal verification approach. Some model checkers accept timed (e.g. \Uppaal{}\footnote{\url{http://uppaal.org/}}) or probabilistic specifications (e.g. PRISM\footnote{\url{http://www.prismmodelchecker.org/}}). Variants of this approach, called \textit{program} model checkers (e.g. Java PathFinder\footnote{\url{https://github.com/javapathfinder}}), operate on the program itself~\cite{Visser2003}.


An example of model checking in the literature is Webster et al.~\cite{webster2011formal, Webster2014}, who use a program model checker to show that an autonomous system controlling a pilotless aircraft meets the requirements for human pilots. The autonomous system is implemented as a \gls{bdi} agent, so its decision making can be checked using the program model checker AJPF~\cite{Dennis2012}. The system is verified against the requirements for human pilots, formalised from the UK Civil Aviation Authority's Rules of the Air. This can be seen as analogous to checking that a human pilot obeys the Rules of the Air. 


Model checking has some advantages over other approaches: model checkers are automatic, which makes them relatively easy to use; also, the concept of checking every state in a model to see if a required property holds is relatively intuitive. However, because model checking exhaustively explores a specification, care must be taken when writing the input specification and choosing properties, to avoid state space explosion (where the number of states that the model checker has to search becomes intractably large). State space explosion, and building the specifications themselves\cite{Rozier2016}, are the two obvious overheads of this approach.

\subsection{Theorem Proving} 
\label{sec:theoremProving}

Theorem Proving is an approach for producing formal proofs of the correctness of a software system. Like model checking, there are a variety of logical systems available and the user is often aided by automatic or semi-automatic theorem proving tools. Formal proofs have the advantage of being able to describe systems with an infinite state space. 

In the literature, Mitsch et al.~\cite{mitsch2013provably} use theorem proving to verify that a ground rover will not collide with, and maintains a sufficient distance from, both stationary and moving obstacles in the rover's environment. Their robot is navigating its environment using the dynamic window algorithm. Both the robot and its environment are modelled in differential dynamic logic (dL)~\cite{Platzer2007}, which is designed for hybrid systems -- that is, systems with both discrete and continuous elements. They report that the theorem prover they used performed 85\% of the proof steps automatically.

Farrell et al.\cite{marie_farrell_formal_2020} use theorem proving to verify a grasping algorithm used by an autonomous robotic arm. The arm is attached to a satellite that is collecting orbital debris. They use the Dafny program verifier to model the algorithm and verify three safety properties about the algorithm. Dafny uses the Z3 automated theorem prover to perform  the checks automatically.

Theorem proving is effective and powerful, but the learning curve for the approach and its tools may be higher than the other approaches mentioned in this section. Also, the concept of theorem proving and the results it produces may be more difficult to explain to stakeholders who do not have a mathematics or formal methods background.

\subsection{Runtime Verification} 
\label{sec:rv}


\gls{rv} uses a monitor, which consumes events or variable updates from a system and compares them to a formal specification of the expected behaviour. If the system’s behaviour differs from that described by the specification, then the monitor can take some mitigating action. \gls{rv} can work while the system is running (online \gls{rv}) or can consume logs of previous runs (offline \gls{rv}).

The mitigating actions that can be implemented depend on the use case and system's architecture. For a simple application of runtime monitoring, the mitigating action could  be to log the failure, or to alert a responsible human. More sophisticated approaches can move towards runtime \textit{enforcement}, where the system is prevented from behaving in a way that does not conform to the expected behaviour. 

One approach in the literature uses a model checker at design-time to verify that a formal specification of an autonomous vehicle in its environment always maintains a safe distance from other vehicles~\cite{aniculaesei2016towards}. When the system is operating, the environment is monitored and if the environment doesn't conform to the design-time model (for example, if there is a vehicle closer than the safe distance) then the system enters a safe stop mode.

\gls{rv} has several benefits. First, it bridges the \textit{reality gap} -- the gap between design-time models and the real world at runtime. This means that \gls{rv} is useful when the system's intended environment becomes impossible to model with enough accuracy; the assumptions and abstractions can be checked as the system is operating. 

Second, \gls{rv} monitors are usually simpler than the system that they are monitoring, which can help when analysing the software element of a \gls{rv} tool. Further, the expected behaviour is a formal specification, so it can usually be verified separately, to ensure that the specified expected behaviour is correct.

Third, \gls{rv} can often reuse an existing formal specification, which enables a workflow where a system is specified, statically verified, and then monitored during its operation~\cite{ferrando_early_2019}. This reuse is especially helpful since building specifications is the bottleneck in the application of most \gls{fm}~\cite{Rozier2016}. 

One obvious downside is that running a monitor uses extra memory, CPU time, and possibly network bandwidth (compared to running the system without a monitor). \textit{Predictive} Runtime Verification~\cite{Zhang2012, Pinisetty2017} aims to reduce the overheads of \gls{rv}, by using specifications of the system's possible behaviour to predict when a property will remain satisfied, which allows the monitor to be paused or stopped. Predictive \gls{rv} also provides a route to runtime enforcement by highlighting the possibility of a violation before it has occurred, so that the violation can be prevented.

\subsection{Formal Synthesis}
\label{sec:synthesis}

Formal Synthesis is an approach that automatically derives low-level controllers for autonomous systems, from high-level task specifications. When used for autonomous robotic systems, the synthesised controllers often control the robot's movement. The utility of this approach lies in automating the conversion of complex task specifications into controllers, especially when they cannot be trivially converted to a sequence of ``go here" statements.

Several studies have shown the utility of formally synthesising (usually movement) plans that satisfy some specified properties (e.g \cite{Loizou2004}). Some have tackled generating movement plans for autonomous robots~\cite{karaman2009sampling, kress2011correct}. Formal synthesis has the potential to be used during execution by the autonomous system itself~\cite{karaman2009sampling}. Since this is a similar technique to model checking, it too can suffer from state space explosion. The examples in both~\cite{karaman2009sampling} and \cite{kress2011correct} plan a short distance ahead to help sidestep the problem of state space explosion. Formal synthesis has also been used to generate controllers for autonomous robots in a team, from a top-down specification of the team's required behaviour~\cite{chen2012formal, chen2013formal}.

Formal synthesis is an active research area, and can be a powerful technique for deriving a controller that implements a particular task or behaves according to certain rules. Various synthesis approaches adapt model-checking algorithms, but for more complex systems this can also cause state space explosion~\cite{chen2013formal}. Automating controller synthesis moves the development overhead to an earlier development lifecycle stage; like with model checking, designing the right specification is key.

\section{Formal Verification Recipes}

\label{sec:verificationTactics}

This section describes five recipes for formally verifying different components or aspects of an autonomous system. As previously mentioned, each recipe makes use of formal approach(es), like those mentioned in the \nameref{sec:approaches} section. The recipes aim to overcome some of the verification challenges that autonomous systems present, and illustrate why \gls{fm} are a set of powerful techniques that should included in the verification repertoire for autonomous systems.

The software in an autonomous system is responsible for making decisions, which brings both new challenges and new opportunities. If an autonomous system is implemented using \gls{ai} that is amenable to inspection (see the \nameref{sec:TypesofAutonomy} section) then this enables us to look inside the `brain' of the system and ensure that it is making correct, ethical, and safe decisions. 

While this paper often focusses on autonomous robotic systems, because of their safety implications, the majority of the recipes that we discuss in this section can be applied to autonomous systems in general. 
The first two recipes, \nameref{sec:execDecisions} and \nameref{sec:controllingMachineLearning}, are directly focussed on the \gls{ai} components making the system's decisions. 
The next two recipes, \nameref{sec:rvSafetyDocs} and \nameref{sec:verifyingMissions}, discuss formal verification of high-level system properties or descriptions (respectively, safety claims and tasks/missions). 
The final recipe, \nameref{sec:logicalBarriers}, is most obvious applicable to autonomous \textit{robotic} systems, but even so this could be applied to general autonomous systems that have decisions that the system \textit{can} but \textit{should not} make.

Autonomous systems often comprise some components that are amenable to \gls{fm} and some that are not~\cite{luckcuck_summary_2019}. Components that are easier to formally verify include an agent that is making the system's executive decisions, like in Cardoso et al.'s example application~\cite{cardoso_towards_2020}. Components that are less easy to apply \gls{fm} to include machine learning components. Focussing the use of \gls{fm} on the components of a system that are amenable to formalisation is useful~\cite{cardoso_heterogeneous_2020, cardoso_towards_2020}, especially if the components that are formalised are those most critical to the safety of the system.
Applying \gls{fm} in only some phases of the system's development is also beneficial, for example formally specifying the system's requirements clarifies its expected behaviour and can also provide an unambiguous starting point for informal verification approaches~\cite{bourbouh_integrating_2021}.

Dynamic verification of autonomous systems should be used in addition to static verification techniques, especially where an autonomous system may encounter humans. This has the extra benefit of helping to bridge the reality gap. This can be implemented as some kind of \gls{rv}, preferably while the system is running (online \gls{rv}) to ensure that the system is still fulfilling its original specification. This is especially useful if the system learns online, is capable of reconfiguring itself, or will be in use autonomously for a long time.
 
Driverless cars are a good example of where both static and dynamic verification would prove useful. We have had many decades of incrementally improving the safety of the physical aspects of cars; the challenge with \textit{autonomous} cars lies in the software that controls the vehicle. When a person learns to drive a car, they must (usually) pass some tests to obtain a licence; the skills that a human needs to display to pass their driving test are also (some of) the requirements of an autonomous system controlling a car. These requirements could be formalised and the autonomous component can be statically checked to ensure that it never chooses to disobey them (unless a human driver would be expected to disobey some rules in a particular situation). Dynamic checks could then be used, against the same formal specification, to show that the driverless car is still displaying the properties needed to get a licence. Interestingly, this could allow a driverless car's `licence' to be revoked if the vehicle is no longer obeying its requirements. This could be caused by a fault in the software or by a mismatch between the system's internal models (defined at design-time) and the real world that the system encounters at runtime -- the \textit{reality gap}. 

While there are many aspects of autonomous software that provide unique challenges, there are other aspects that resemble  problems similar to those in other types of software system. Approaches to verifying these similar problems can often be reused or adapted for autonomous systems. 

For example, Multi-Agent Systems (MAS) and robot swarms exhibit challenges that are similar to multi-threaded systems, such as scheduling and resource allocation. Other autonomous systems choose which tasks to complete within a particular time frame (such as in~\cite{Hawes2017}), which makes the similarity to scheduling more obvious. 

\gls{ros} is often used in the literature for implementing autonomous robotic systems, and \gls{fm} has been used to enable the verification of its producer-consumer communication mechanism~\cite{halder_formal_2017}. \gls{ros} systems are composed of nodes that communicate over a network. Each node can publish data to named channels, and subscribe to receive data from channels. Nodes must state the rate at which they will publish to a channel, and the size of buffer used when subscribing to a channel. A formal model is used to capture the publishing rate and buffer sizes of the \gls{ros} nodes, for a particular program. A model checker is used to verify that the buffers never overflow. This is an important property to verify, and this technique uses an abstraction to produce a model that captures just the pertinent details.

The remainder of this section describes the five verification recipes and the utility of \gls{fm} approaches in applying the recipes. The five recipes in this section are \nameref{sec:execDecisions}, \nameref{sec:controllingMachineLearning}, \nameref{sec:rvSafetyDocs}, \nameref{sec:verifyingMissions}, and \nameref{sec:logicalBarriers}. While this is not an exhaustive list of where and how \gls{fm} approaches can be applied to autonomous systems, this is the starting point for a larger, cookbook of formal verification recipes.

\subsection{Recipe 1: Verifying Executive Decisions}
\label{sec:execDecisions}

The executive decisions of an autonomous system should be taken by rational, symbolic \gls{ai} components (for example \gls{bdi} agents~\cite{RaoG95}) that enable the verification of the choices being made.
Using this type of autonomy to make the system's executive decisions provides the opportunity to inspect and verify those decisions, something that is currently almost impossible to do with human decision-making. This opportunity should not be wasted, and is another instance where autonomous systems should be designed for verification. 

With verifiable autonomy, the system's choices can be checked to ensure that they obey safety requirements. This could be seen as akin to being able to check that a new driver will obey all the relevant driving laws before they get their driving licence.

Specifically in the case of \gls{bdi} languages, the program model checker AJPF~\cite{Dennis2012} can directly verify properties about the agent program. An example of this approach can be found in~\cite{Webster2014}, where the authors encode the UK Civil Aviation Authority's Rules of the Air and verify that the \gls{bdi} agent making the executive decisions always abides by these rules. This example is similar to \nameref{sec:rvSafetyDocs}.

Using this recipe requires autonomy that is amenable to inspection and verification. 
If \gls{ai} techniques that are opaque and less amenable to inspection and verification are used, then the dynamic verification described in \nameref{sec:rvSafetyDocs} may be useful instead. However, even if the static verification described in this recipe is used, the checks assume the correctness of the system's sensors. Given the possibility of sensor failure or incorrect interpretation of noisy sensor data, both of these verification techniques could be used alongside each other, for extra assurance. 

The flexibility of this approach has also been used to check that autonomous systems are making ethical decisions. This is an open challenge, but must be dealt with if autonomous systems are to be trustworthy. Often, an autonomous system's choices are evaluated as being either ethical or not~\cite{Arkin2009, Arkin2012, Brutzman2013}. This is a useful starting point, but real-life situations are unlikely to be so simple. An example of taking a less dichotomous approach is~\cite{Dennis2016a}, which describes an autonomous system (using a \gls{bdi} language) that can reason about the ethical weighting of its choices. These programs are directly verified to check more complex ethical properties, such as the system choosing the least-worst option in situations where only unethical choices are  available.

\subsection{Recipe 2: Controlling Machine Learning}
\label{sec:controllingMachineLearning}

A machine learning component should be paired with a \textit{governor} that ensures that the output of the machine learning component remains within a correct operating envelope -- compare with mechanical governors\cite{maxwell_i_1868}. Crucially, this requires an architecture where the governor is able to prevent an incorrect decision from the machine learning component from propagating to the rest of the system. This recipe is useful where incorrect output from a machine learning component could cause the executive control component(s) to make a decision that leads to a failure, for example a vision classifier. 

Machine learning can be useful for discovering solutions to problems where there is little or no prior knowledge of the context; for example, classifying objects or exploring unknown environments. But ensuring that machine learning remains within a safe operating envelope can be challenging~\cite{mallozzi_keeping_2018}. For example, as raised by Koopman and 
Wagner, and summarised in~\cite{mallozzi_keeping_2018}; machine learning components can suffer from: 
\begin{itemize}
\item \textbf{Over-fitting}, where the machine learning component too closely corresponds to the training data and therefore might not be reliable on real-world data;
\item \textbf{Black Swan Events}, which are special cases that the component has not seen in the training data and therefore might not react to correctly; and,
\item \textbf{Reward Hacking}\cite{everitt_reinforcement_2017}, where an incorrectly specified reward function might be exploited by the machine learning component to gain rewards for unsafe actions or predictions. 
\end{itemize}
\noindent 
Machine learning components should be separated from the executive control of the system, and not linked directly to actuation. This recipe enables verification and monitoring techniques to be used to mitigate the risk of incorrect output from a machine learning component causing the overall system to perform unsafe behaviour. 

A governor could be implemented as a purely software monitor, but this just shifts the burden of verification from the autonomous system to the software monitor, which is likely to be simpler but will face similar verification challenges. Instead, a formal runtime \textit{enforcement} approach adds confidence to the monitoring used by the governor and can be directly verifiable itself. It also has the benefit that the specification for the correct operating envelope is unambiguous and provably correct. 

Governors used in this way could enforce relatively simple properties, like a confidence threshold for a machine learning verdict. But they could also track the ongoing output of the machine learning component to enforce more complex properties. As an example, consider a driverless car that uses a machine learning classifier to identify obstacles. The classifier might be reliable on clear distinctions between cars, bicycles, and pedestrians. However, it might not have been trained on `messy' scenes showing, for example, a pedestrian pushing a bicycle across the road. In this situation the classifier might differ between a classification of `pedestrian' and `bicycle'. A governor introduced here could enforce a property saying: if the classifier hasn't settled on a result within a given time frame, then the vehicle would slow down and prepare to stop. This would enforce the safe behaviour expected of a human driver.

This recipe can be found in the literature. The \textsc{WiseML} approach~\cite{mallozzi_runtime_2019} is an \gls{rv} framework that enforces invariants in autonomous systems that use Reinforcement Learning (RL). It is demonstrated on an autonomous system that is using an RL algorithm to explore an unknown environment. Another implementation route, for systems using the \gls{ros} middleware, is ROSMonitoring\footnote{\url{https://github.com/autonomy-and-verification-uol/ROSMonitoring}}~\cite{Ferrando20a}, which is an \gls{rv} framework that can drop messages that do not conform to a formal specification. This could be used to vet the output from the machine learning component and prevent incorrect messages from reaching the rest of the system. Finally, the concept of \textit{shields}~\cite{
konighofer_shield_2017, alshiekh_safe_2018, konighofer_online_2021} is another route. A shield is a correct-by-construction runtime enforcer, synthesised from a safety property. Like the idea behind ROSMonitoring, the shield operates at runtime to detect and overwrite violations of the safety property, so that the system always obeys its safety requirements.
 
\subsection{Recipe 3: Verifying and Enforcing Safety Claims}
\label{sec:rvSafetyDocs}

\gls{rv} techniques provide a good way of showing that a system's safety claims are fulfilled at runtime, and can be extended to enforce the safety claims too. Static verification is useful (for example, \nameref{sec:execDecisions} could also be used here) but dynamic checking and enforcement of safety requirements provides strong evidence for the system obeying its requirements in more realistic settings. This kind of evidence is especially important where a system needs regulator approval before operation. 

The safety claims or properties in a safety document describe how the system's behaviour is constrained to maintain safety. Examples include keeping a minimum distance from obstacles, limiting the speed or force that the system can apply, or constraining a behaviour to a certain location (such as a laser cutter only being activated when it is pointing towards a safe cutting zone).

As described in the \nameref{sec:rv} section, \gls{rv} monitors the running system and compares its behaviour to a formal specification of its expected behaviour. Using the system's safety documents as the source for the specification of the expected behaviour enables these safety claims to be checked and enforced. 

In previous work, we formalise the safety subsystem described in a safety design document and use the resulting specification as a runtime monitor~\cite{luckcuck2020monitoring}. The specification is checked against the safety requirements (also extracted from the safety document) to validate the model. This has the side-benefit of checking the design of the safety subsystem. This specification can also be used to guide the implementation of the safety subsystem, if it is still being developed.

The monitoring provided by \gls{rv} could be extended to enable enforcement of the safety requirements. One approach to runtime enforcement could be to use \textit{predictive} \gls{rv}\cite{Zhang2012} to try to foresee the future violation of a safety property and use this information to prevent the violation. The workflow described in~\cite{luckcuck2020monitoring} could be used to guide the formalisation of existing safety documentation to use as the predictive model. The benefit of predictive \gls{rv} is that it might also foresee the continued satisfaction of a property, which means that the monitor could be switched off to save CPU and memory usage. 

Similarly to \nameref{sec:controllingMachineLearning}, runtime enforcement requires a system architecture that allows the monitor to intercept instructions that would cause the system to violate a safety requirement. An example of how to achieve this is presented in~\cite{Ferrando20a}, where a monitor is placed inside a \gls{ros} system and is able to drop messages that do not conform to its specification of acceptable behaviour. This means that the unsafe messages do not get acted on, so the safety requirements are enforced at runtime.

\subsection{Recipe 4: Verifying Task and Missions}
\label{sec:verifyingMissions}

Autonomous systems are often given goals to complete, to guide their behaviour, which could be viewed as high-level programs. This means that the arrangement of goals can be a source of failure and so they are a target for verification. 

Often, these goals and how to achieve them are described as \textit{tasks} or \textit{missions}. Here, we will use 
\textit{task} to mean a collection of actions, such as moving to a waypoint or inspecting an object; and \textit{mission} to mean a collection of tasks to achieve some goal, such as inspecting every waypoint in the map. For some systems, missions and tasks might be set by a human user or operator, in others the system might make more decisions about its mission or how to achieve a goal. 

The tasks should be identified and well defined, this makes the system's capabilities clear to both the implementers and the users of the system. This should include, for example: the task's context and inputs (and any other pre-conditions), how it performs the task, the expected outputs (and any other postconditions), what might cause the task to fail and how the system should recover, etc. 

The software implementation of a task or a mission is often written in a general purpose programming language, because it must be executed by the software controlling the system. This means that verification approaches (especially formal verification approaches) could be applied directly to the software. However, there is still utility in taking a formal approach to the design of tasks and missions, similarly to the benefit of formalising a system's requirements. 

Using informal notations (such as flowcharts or communication diagrams) is useful for clarifying the intended behaviour of tasks and missions, and would enable formalisation at a later stage. For someone trying to formalise a system, this would already be an improvement over trying to extract a formal specification from program code or natural-language documentation.

Formalising the system state that the task expects before it starts (pre-conditions) and how the task will have changed the system state after it has finished (post-conditions) is a good start for formalising the design of a task -- as in Hoare triples~\cite{hoare_axiomatic_1969}. These formal conditions clarify the task's requirements and can be used to verify that its implementation meets these requirements (either statically or by using \gls{rv}). This lightweight formal approach could identify where a mission is incorrectly constructed. For example, it could highlight where one task's pre-condition relies on something that the previous task's post-condition does not guarantee. This design-time check would enable developers to fix a potential failure before implementation.

When specifying a task it is important to remember that the system will interact with a (virtual or physical) environment, so an action might fail or otherwise not behave as intended.
This could be due to a changing environment, unforeseen conditions, or a hardware failure. Once a failed action has been detected, the system can take steps to recover. This should also be specified in the task's design, including a fallback task (or tasks) to safely recover from the failure or move to some known safe state. For example the mission specification language Sophrosyne enables the specification of tasks and fallback actions for if the task fails~\cite{viard_mission_2020}).
Another example is the extensions to agent programming languages that enable the description of tasks (`plans' in the agent program) that can (attempt) to reconfigure themselves if they fail~\cite{dennis_actions_2014, stringer_adaptable_2020}. This includes describing the pre- and post-conditions for each task and then using an automated planner to attempt to repair the plan.
    
The mission pattern catalogue described in~\cite{menghi_specification_2019} provides an example of how to formally verify missions. The pattern catalogue is a taxonomy of mission categories -- including avoidance, surveillance, and coverage patterns -- created from a survey of the robotic literature. Each category is populated by several patterns, and each pattern is described in English and in temporal logic. Their tool, {PsALM}~\cite{menghi_psalm_2019}, enables missions to be constructed from the patterns in the taxonomy, from which a temporal logic specification of the mission is synthesised. This enables missions to be formally verified before operation. 

\subsection{Recipe 5: Logical Barriers}
\label{sec:logicalBarriers}

Autonomous robotic systems may have a physical barrier to contain them within a safe area, this could be complemented by a formally verifiable \textit{logical} barrier. A logical barrier is a software restriction on the area in the  operating environment that an autonomous robotic system can move to. More generally, a logical barrier could be a restriction on the decisions that an autonomous system can make but should not, for legal, ethical, or safety reasons.

The concept of a logical barrier is not entirely new, for example it can be seen in autonomous vehicle platooning, where software controllers ensure that vehicles keep a safe distance from the vehicle in front of them~\cite{colin_using_2009, kamali_formal_2017,karoui_dual_2017}; and where mobile robots need to avoid obstacles, for example in~\cite{mitsch2013provably, aniculaesei2016towards}, or stay within a predefined area (geofencing)~\cite{dill_safeguard_2016, viard_mission_2020}. It is similar in intent to the work on barrier certificates~\cite{goos_safety_2004}, which is a technique for showing that a hybrid system does not enter an unsafe region.

This concept has obvious applications where autonomous robotic systems might be close to humans, who must be protected from physical harm. In this situation, a physical barrier might be a wall or other buffer that stops the system moving too far from one position or too close to other objects or people. For example, a robotic arm surrounded by a safety barrier that stops it from moving outside of a safe working area. 

Using a logical barrier provides a penultimate line of defence; the software will stop the system from getting close to the physical barrier, but the physical barrier is there in case the software safety system fails. The logical barrier could be in the same place as the physical barrier or it could be slightly offset to provide a logical buffer zone, for example to allow enough space for the robotic system to slow down before hitting the physical barrier. The placement of a logical barrier should be based on an assessment of concerns like robotics systems' maximum speed or the force that it can deliver.

One approach to implementing this recipe is to use \gls{rv} to provide runtime enforcement. So long as the monitor can listen to the system's behaviour, this approach would not require a particular type of autonomy. 
However, the architecture of the system must be arranged so as to allow the monitor to change the system's behaviour, to prevent it from crossing the logical barrier.
As previously mentioned, \gls{rv} can add computational overheads, so it must be implemented carefully to avoid potentially delaying the system.

Another approach to implementing this recipe could be to incorporate the logical barriers into the executive decision making component, in such a way that it is formally verifiable. This is related to \nameref{sec:execDecisions}, so there is a similar requirement that the high-level decision making uses \gls{ai} techniques that can be inspected (symbolic \gls{ai}). This enables the verification that, for example, the system never chooses to move past the logical barrier. 

Using the \gls{rv} approach checks the dynamic behaviour of the system; checking the system's decision-making is static verification. Even if the system never \textit{chooses} to move past the barrier, it could conceivably do so by accident. Clearly, a combination of these two approaches would work well.

The need for logical barriers, when the physical barriers exist, is possibly subtle; but they provide an extra layer of assurance. Logical barriers could also be useful in situations where a physical barrier is desired, but not possible -- though, the combination of both types of barrier would be the most effective. This could be helpful in establishing enough confidence in an autonomous system's safety to enable field tests; without such confidence, the risk of injury during field tests might be considered too high.

\section{Conclusion}
\label{sec:conclusion}

This paper answers the question \textit{Why use Formal Methods for Autonomous Systems?} by describing the utility of four often-used formal verification approaches and by presenting five formal verification recipes for autonomous systems. The formal approaches explain what \gls{fm} can do, and the recipes describe how they can be used to verify particular components or aspects of an autonomous system.

\gls{fm} are a powerful technique to have in one's repertoire when verifying autonomous systems. They can be used in many stages of the systems development life cycle, providing both static and dynamic verification; and they enable exhaustive, automatic checks over unambiguously-defined specifications. Despite persistent myths about \gls{fm}~\cite{Hall1990,Bowen1995a}, they have been used in industry~\cite{Woodcock2009} and in the academic literature on formally verifying autonomous systems~\cite{luckcuck_formal_2019}. 

Applying formal verification needs a good description of the system and its requirements, just like any other verification technique. Improving these descriptions requires better collaboration between \gls{fm} practitioners and developers of autonomous systems. Formal verification approaches need to be able to cope with the architectures, requirements, and challenges of autonomous systems. In turn, autonomous systems engineering must ensure that it can produce systems that are amenable to robust verification techniques and provides clear definitions of the system's scope and requirements. The recipes presented here are grounded in our previous survey~\cite{luckcuck_formal_2019} and research experience\cite{fisher_overview_2021} of formal verification for safety-critical autonomous systems.

As previously mentioned, some of the recipes describe multiple ways that they can be implemented. It is also possible that more than one of the recipes will be applied to the same system at the same time, and some of the recipes suggest how they can be used with another.  In previous work~\cite{Farrell2018}, we argue that autonomous systems (particularly those embedded in robotic systems) need an integrated approach to verification, because of their complexity and blending of different specialist fields into one system. This means that it is likely that a system will need a combination of formal and informal verification approaches. One methodology for combining different approaches is the \textit{Corroborative \gls{vnv}} approach ~\cite{websterCorroborative2020}, which describes how to use formal and informal models of the same system to support wholistic verification. We consider this to be an \textit{integrated} approach, because it combines (or integrates) several approaches.

The formal verification recipes discussed in this paper are a non-exhaustive snapshot of the utility of formal verification, but they are the starting point for a larger cookbook of recipes. 
There is currently little guidance about which \gls{fm} to use, and best practice on how to use them. Efforts like the collection of relevant surveys and evaluations of the state-of-the-art in \gls{fm} by the Formal Methods Europe association\footnote{Formal Methods Europe `Choosing a Formal Method': \url{https://fmeurope.org/choosingaformalmethod/}} are a good starting point.
The recipes we present in this paper kickstart the discussion about systematic ways of formally verifying aspects of autonomous systems. Useful future work in this area includes implementing the recipes presented here to refine them, and analysing the literature and case studies for other useful formal verification recipes.

Because of the safety and security implications of an autonomous system failing, the most robust methods should be chosen for each system component and development phase. 
Formal methods are, by no means, a panacea for the challenges of engineering safe, correct, and trustworthy autonomous systems. However, the recipes presented in this paper illustrate why \gls{fm} are useful for the verification of autonomous systems -- alongside other verification techniques, including software testing, simulation-based testing, or physical tests. \gls{fm} can provide robust specification and verification of many components and aspects of an autonomous system, so formal methods should be included in the repertoire of verification techniques used in the engineering of autonomous systems.

\bibliographystyle{plain}

\bibliography{iwassJournalPaper.bib} 

\end{document}